\documentclass[12pt]{iopart}

\usepackage{iopams} 
\usepackage{bm,graphics,graphicx,epsfig,color}

\def\apj{{\em Astrophys.\ J.\ }}
\def\apjl{{\em Astrophys.\ J.\ Lett.\ }}
\def\prd{{\em Phys.\ Rev.\ D }}
\def\cqg{{\em Class.\ Quantum Grav.\ }}

\begin{document}

\title[Testing no-hair theorems]{Testing the black hole no-hair theorem at the galactic center: Perturbing effects of stars in the surrounding cluster}

\author{Laleh Sadeghian$^{1}$ and Clifford M. Will$^{1,2}$}

\ead{lsadeghian@physics.wustl.edu, cmw@wuphys.wustl.edu}
\address{
$^1$ McDonnell Center for the Space Sciences, Department of Physics,
Washington University, St. Louis MO 63130 USA
 \\
$^2$ GReCO, Institut d'Astrophysique de Paris, UMR 7095-CNRS,
Universit\'e Pierre et Marie Curie, 98$^{bis}$ Bd. Arago, 75014 Paris, France
}

\begin{abstract}
Observations of the precessing orbits of stars very near the massive black hole in the galactic center could provide measurements of the spin and quadrupole moment of the hole and thereby test the no-hair theorems of general relativity.   Since the galactic center is likely to be populated by a distribution of stars and small black holes, their gravitational interactions will perturb the orbit of any given star.  We estimate the effects of such perturbations using analytic orbital perturbation theory, and show that for a range of possible stellar distributions, and for an observed star sufficiently close to the black hole, the relativistic spin and quadrupole effects will be larger than the effects of stellar cluster perturbations.  Our results are consistent those from recent numerical $N$-body simulations by Merritt {\em et al.}.

\end{abstract}

\maketitle

\section{Introduction and summary}
\label{sec:intro}

The center of our galaxy has become an active arena for studying possible tests of general relativity (GR) in the strong-field regime, because of the near certainty that it harbors a 4 million solar-mass black hole, colloquially denoted SgrA* (see \cite{alexander05,genzel10} for reviews).  Numerous authors have studied the observability of  relativistic effects in the vicinity of the black hole, in the observable motion and behavior of orbiting stars 
\cite{jaroszynski,fragile,rubilar,weinberg,zucker,kraniotis,angelil1,angelil2,angelil3}, 
in the effects of lensing \cite{binnun}, or in the properties of accretion phenomena \cite{psaltis1,psaltis2,psaltis3}.  

One of us recently suggested that observations of a hypothetical class of stars orbiting very close to the galactic center black hole could provide tests of the so-called ``no-hair'' theorem of general relativity \cite{willnohair}.  Specifically, measurements of the precessions of the orbital planes of a number of stars with an accuracy of $10$ microarcseconds ($\mu$as) per year could determine both the angular momentum $J$ and the quadrupole moment $Q$ of the black hole, and thereby test the constraint $Q = - J^2/Mc^2$ imposed by the Kerr solution of general relativity.  Detection of such stars and achieving the required astrometric accuracy are goals of the next-generation of near-infrared, adaptive optics interferometry being designed and built by a number of groups \cite{gillessen,eisner}.

However, in assessing the feasibility of such strong-field GR tests, one must inevitably address potential complications, notably the perturbing effect of the other stars that may also reside in a cluster close to the black hole.  Using $N$-body simulations, Merritt {\em et al} (\cite{mamw}, hereafter referred to as MAMW) showed that for a range of possible stellar and stellar-mass black hole distributions within the central few milliparsecs (mpc) of the black hole, there could exist stars in eccentric orbits with semi-major axes less than $0.2$ milliparsecs for which the orbital-plane precessions induced by the stars and black holes would not exceed the relativistic precessions.   These conclusions were gleaned from thousands of simulations of clusters ranging from seven to 180 stars and stellar mass black holes orbiting a $4 \times 10^6 \, M_\odot$ maximally rotating black hole, taking into account the long-term evolution of the system as influenced by close stellar encounters, dynamical relaxation effects, and capture of stars by the black hole.

In this paper we study the extent to which the conclusions of MAMW can be understood, at least within an order of magnitude, using analytic orbit perturbation theory.  We calculate the average change in the orientation of the orbital plane of a given ``target'' star orbiting the massive black hole, as determined by its inclination and ascending node angles $i$  and $\Omega$, induced by the Newtonian gravitational attraction of a distant third star (which could be either inside or outside the target star's orbit).
The perturbing accelerations are expanded in terms of multipoles through $\ell = 3$.   We then calculate the root-mean-square variation of each orbit element, averaged over all possible orientations of the perturbing star's orbit, and averaged over a distribution of orbits in semi-major axis and eccentricity, arguing that this will give an estimate of the ``noise'' induced by the graininess of the otherwise spherically symmetric perturbing environment.

\begin{figure}[t]

\includegraphics[width=5in,angle=270]{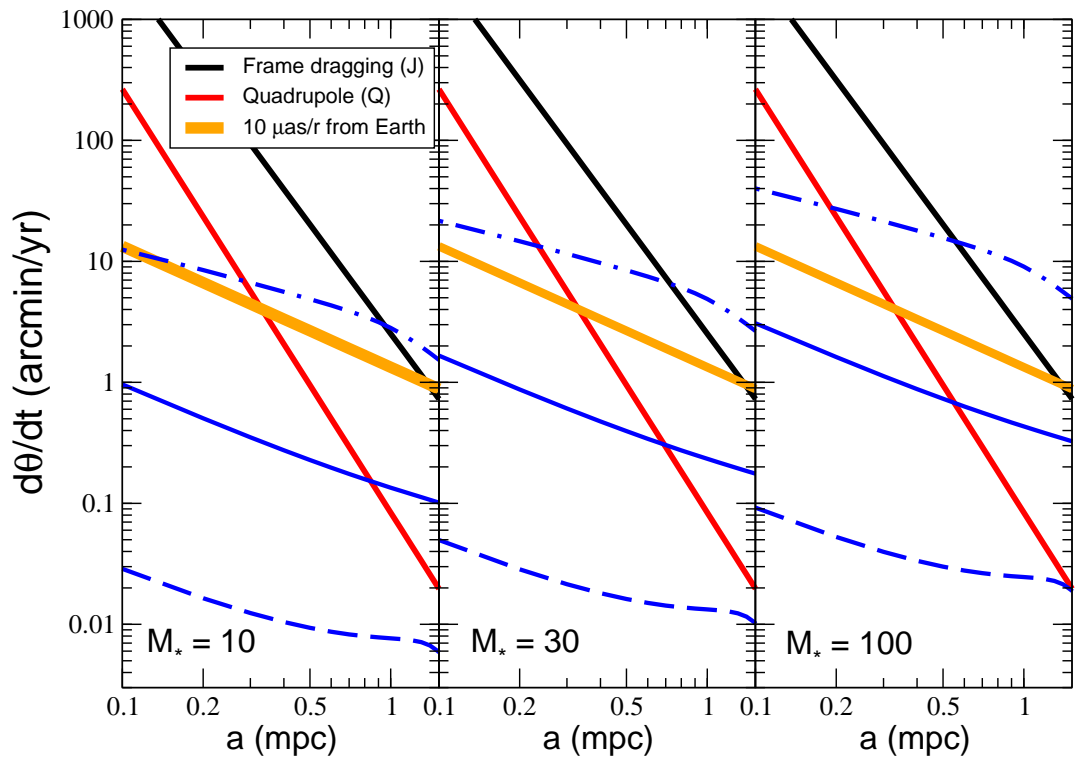}

\caption{R.m.s.\ precession $d\theta/dt =  (\langle \dot{\imath}^2 \rangle + \sin^2 i \langle \dot{\Omega}^2 \rangle)^{1/2}$ for a target star with $e=0.95$ plotted against semi-major axis, for three models with $\gamma =2$, $\beta = 0$, $R = 1$.  $M_\star$ denotes the total mass within one mpc.    Shown (blue in color version) are results from
Integration I (dashed curves), Integration II (solid curves) and Integration III (dot-dash curves).  Also shown are the amplitudes of frame-dragging (black in color version) and quadrupole (red in color version) relativistic precessions for the corresponding star, assuming a maximally rotating black hole.  Wide line (orange in color version) denotes the precession corresponding to an observed astrometric displacement of 10 $\mu$as/yr. }

\label{fig1}
\end{figure}

Figure 1 shows the results for three stellar distribution models, among the set of models studied by MAMW.  They  have number densities that vary as $1/a'^{\gamma}$, where $a'$ is the semi-major axis of the perturbing star, and have eccentricity distributions that vary as $(1-e'^2)^{-\beta}$; in Fig. 1, $\gamma =2$ and $\beta = 0$, corresponding to a distribution with isotropic velocity dispersion.  They have an equal number of $1M_\odot$ stars and $10M_\odot$ black holes.  The label $M_\star$, chosen to parallel the notation of \cite{mamw}, denotes the total mass within one mpc of the black hole; the three cases correspond to a total number of perturbing bodies within a radius of four mpc of 7, 21 and 72, respectively.  The target star has eccentricity $e = 0.95$, and its semi-major axis $a$ ranges from $0.1$ to $2$ mpc.   Plotted is the rate of precession of the vector perpendicular to the orbital plane,  $d\theta/dt$, observed at the source, in arcminutes per year, calculated using three ways of carrying out the integrals over the stellar distribution.   The dashed line denotes an integration (I) in which all perturbing stars are assumed to be sufficiently far from the target star at all times that their pericenters are outside its apocenter or that their apocenters are inside its pericenter.  The solid line denotes an integration (II) in which closer encounters are permitted, limited by demanding that all perturbing stars be on orbits such that the higher $\ell$ contributions to $d\theta/dt$ be at worst comparable to the contribution at lowest order in $\ell$.  The dot-dashed line denotes an integration (III) which uses a fitting formula that interpolates between the extreme limits of a perturbing star well outside the target star, and a perturbing star well inside the target star; in this case the integration is over the entire stellar distribution.  The orange band in each panel denotes the value of $d\theta/dt$ corresponding to an astrometric precession rate $d\Theta/dt$ of $10 \, \mu$as per year as seen from Earth, given by 
\begin{equation}
\frac{(d\theta/dt)_{\rm source}}{({\rm arcmin/yr})}
\approx  \frac{1.3}{\tilde a} \frac{(d\Theta/dt)_{\rm Earth}}{(10 \, \mu {\rm as/yr})} \,,
\end{equation}
where ${\tilde a}$ is the semi-major axis in units of mpc; we use 8 kiloparsecs as the distance to the galactic center.

 Also plotted are the amplitudes of the frame-dragging and quadrupole precessions for a Kerr black hole, given by \cite{willnohair}
\begin{eqnarray}
A_J &=& \frac{4\pi}{P} \chi\left[\frac{GM}{c^2 a(1-e^2)}\right]^{3/2} 
\nonumber \\
& \approx& 0.769 (1-e^2)^{-3/2}\chi {\tilde a}^{-3} \,
{\rm arcmin \, yr^{-1}}  \,,\label{eq:defaj}
\\
A_Q &=&
 \frac{3\pi}{P} \chi^2 \left[ \frac{GM}{c^2 a(1-e^2)}\right]^{2} 
\nonumber \\
& \approx& 7.97 \times 10^{-4} (1-e^2)^{-2}\chi^2 {\tilde a}^{-7/2}
{\rm arcmin \, yr^{-1}}  \,, 
\label{eq:defaq}
\end{eqnarray} 
where $P=2\pi (a^3/GM)^{1/2}$ is the orbital period, and where $\chi = Jc/GM^2$ is the dimen\-sion\-less Kerr spin parameter, set equal to its maximum value of unity in Fig.\ 1.

Because Integration I keeps the stars far from the target star, the precessions are small.  By contrast, the fitting formula of Integration III is large for very close encounters, so not surprisingly, the precessions from that method are large.  Integration II gives results intermediate between the two.  Interestingly, the spread between these methods is roughly consistent with the spread between individual precessions obtained in the $N$-body simulations of MAMW.  This can be seen in the top panel of MAMW, Fig.\ 7, which corresponds to the middle panel of Fig.\ 1 (to properly compare the two figures, one must translate between $d\theta/dt$ and $d\Theta/dt$).   It can also been in the bottom panel of MAMW Fig.\ 5, where the points labelled by $\times$ indicate the mean precessions in the absence of black hole spin, for the same three stellar distributions as are shown in Fig.\ 1.   Thus we regard our three integration methods as giving a reasonable estimate of the range of stellar perturbations.

Comparing the three stellar distributions shown in Fig.\ 1, we see that the effects vary roughly as $N^{1/2} \propto M_\star^{1/2}$, as expected, from the nature of our r.m.s. calculation. 

We consider eight different stellar distribution models, and for seven of them, consider models with equal numbers of stars and black holes, and models with only stars, totalling 15 models.  In all but one case, the precessions are generally smaller than the ones shown in Fig.\ 1, and that case is a centrally condensed model with a non-isotropic velocity dispersion leading to a preponderance of highly eccentric orbits.   We conclude that, for a target star in a very eccentric orbit with $a < 0.2$ mpc, there is a reasonable possibility of seeing relativistic frame-dragging and quadrupole effects above the level of $10 \, \mu$as/yr without undue interference from stellar perturbations.

The remainder of this paper gives the details underlying these results.  In Sec.\ \ref{sec:theory}, we describe the basic orbital perturbation theory leading to the orbit-averaged variations in the orbital elements of a target star.  In Sec.\ \ref{sec:distribution} we calculate the r.m.s. variations of the elements $i$ and $\Omega$ by averaging over distributions of perturbing stars.  Section \ref{sec:numerical} gives the numerical results and compares them with those of MAMW.  Concluding remarks are made in Sec.\ \ref{sec:conclusions}.  Appendix A lists the higher-order contributions to the r.m.s.\ variations, Appendix B derives the minimum distance from the black hole reached by a body that avoids either tidal disruption or direct capture, and Appendix C shows that the effects of tidal deformations on the orbital planes of stellar orbits are negligible.

\section{Orbital perturbation theory at the galactic center}
\label{sec:theory}

\subsection{Basic equations}

In Newtonian theory, 
the acceleration $\bm{a} \equiv \bm{a}_1 - \bm{a}_2$  of a target star with mass $m_1$ relative to a massive black hole (MBH) with mass $m_2$ in the presence of a perturbing star with mass $m_3$ is given by

\begin{equation} 
\label{EOM}
\bm{a}=-\frac{Gm_2\bm{r}_{12}}{r_{12}^3}-\frac{Gm_3\bm{r}_{13}}{r_{13}^3}-\frac{Gm_1\bm{r}_{12}}{r_{12}^3}+\frac{Gm_3\bm{r}_{23}}{r_{23}^3}
\,,
\end{equation}
where  $G$ is Newton's constant, $\bm{r}_{ab} = \bm{r}_{a}-\bm{r}_{b}$, and $r_{ab} = |\bm{r}_{ab}|$.

For a perturbing star inside the orbit of the target star (``internal'' star), with $r_{32} \ll r_{12}$, Eq.\ (\ref{EOM}) can be expanded as
\begin{equation} \label{EOM,in}
{a}^i=-\frac{G(m_1+m_2+m_3){r}^i}{r^3}+\frac{Gm_3{R}^i}{R^3}+Gm_3\sum_{\ell=1}^{\infty}\frac{1}{\ell!} R^L \partial^{\langle iL \rangle} \left (\frac{1}{r} \right ) \,,
\end{equation}
where $\bm{r} \equiv \bm{r}_{12}$ and $\bm{R} \equiv \bm{r}_{23}$; the capitalized superscripts denote multi-indices, so that $R^L \equiv R^i R^j \dots R^{k_\ell}$, and similarly for the partial derivatives; $\langle \dots \rangle$ denotes a symmetric trace-free product. 

For a perturbing star outside the orbit of the target star (``external'' star), with $r_{12}\ll r_{23}$, the expansion takes the form

\begin{equation} \label{EOM,out}
{a}^i=-\frac{G(m_1+m_2){r}^i}{r^3}+Gm_3\sum_{\ell=1}^{\infty}\frac{1}{\ell!} r^L \partial^{<iL>}\left (\frac{1}{R} \right ) \,.
\end{equation}
Because $m_1 \ll m_2$ and $m_3 \ll m_2$, and because we are only concerned in what follows with orbital plane effects, we can replace both $m_1 + m_2$ and $m_1 + m_2 + m_3$ with a single $M$, effectively the mass of the MBH.  

Establishing a reference XY plane and a reference Z direction, one defines the standard ``osculating'' orbital elements.  The inclination $i$ relative to the reference plane, and the angle of ascending node $\Omega$ between the X axis and the line where the orbital and reference planes intersect, fix the orientation of the orbital plane in space.  The semi-major axis $a$, eccentricity $e$, and pericenter angle $\omega$ determine the orbit in the orbital plane.  The true anomaly $f \equiv \phi - \omega$ is measured in the orbital plane from the pericenter to the location of the body. It is also useful to define an auxiliary angle of pericenter $\varpi=\omega+\Omega \cos{i}$ which represents a kind of angle measured from the reference X-direction, rather than from the nodal line.      

The unit vector $\bm{n}$ pointing from the MBH to the target star, and the orthogonal unit vectors $
\bm{\lambda}$ and $\bm{h}$ are given by 
\begin{eqnarray} \label{n}
{\bm n}&=&\bm{e}_X[\cos{(\omega+f)}\cos{\Omega}-\sin{(\omega+f)}\sin{\Omega}\cos{i}]  
\nonumber \\ 
&&  +\bm{e}_Y[\cos{(\omega+f)}\sin{\Omega}+\sin{(\omega+f)}\cos{\Omega}\cos{i}]
\nonumber \\ 
&& +\bm{e}_Z[\sin{(\omega+f)}\sin{i}] \,,
\nonumber \\
 \label{lambda}
\bm{\lambda}&=&-\bm{e}_X[\sin{(\omega+f)}\cos{\Omega}+\cos{(\omega+f)}\sin{\Omega}\cos{i}]  \nonumber \\ 
&& -\bm{e}_Y[\sin{(\omega+f)}\sin{\Omega}-\cos{(\omega+f)}\cos{\Omega}\cos{i}]
\nonumber \\ 
&&+\bm{e}_Z[\cos{(\omega+f)}\sin{i}] \,,
\nonumber \\
\bm{h} &=& \sin i (\bm{e}_X \sin \Omega - \bm{e}_Y \cos \Omega ) + \bm{e}_Z \cos i \,,
\end{eqnarray}
where $\bm{h}$ is normal to the orbital plane.  We also have the osculating orbit definitions $r \equiv p/(1+e \cos f)$, $h \equiv |{\bf r} \times {\bf v} | \equiv (GMp)^{1/2}$, $d\phi/dt \equiv h/r^2$, and $p \equiv a(1-e^2)$ for the target star, and $R \equiv p'/(1+e' \cos F)$, $h' \equiv |{\bf R} \times {\bf V} | \equiv (GMp')^{1/2}$, $d\phi'/dt \equiv h'/{r'}^2$, and $p' \equiv a'(1-{e'}^2)$ for the perturbing star, along with its orbital elements $i'$, $\Omega'$ and $\omega'$.   

Defining the perturbing acceleration to be everything in Eqs.\ (\ref{EOM,in}) and (\ref{EOM,out}) except the leading acceleration $-GM{\bf r}/r^3$, and defining $\mathcal R$, $\mathcal S$, and $\mathcal W$ to be the components of the perturbing acceleration along  $\bm{n}$, $\bm{\lambda}$, and $\bm{h}$ respectively, the Lagrange planetary equations for variations with time of the target star's orbital elements are given by (see, eg. Sec.\ 7.3 of \cite{tegp}), 
\begin{eqnarray}
\label{a}
\frac{da}{dt }&=&\frac{2a^2}{h} \left (\mathcal S \frac{p}{r}+\mathcal R e \sin{f} \right ) \,,
\\ \label{e}
\frac{de}{dt}&=&\frac{1-e^2}{h} \left (\mathcal R a \sin{f}+\frac{\mathcal S}{er}(ap-r^2) \right )\,,
\\  \label {i}
\frac{di}{dt} &=&\mathcal W \frac{r}{h} \cos{(\omega+f)} \,,
\\ \label{Omega}
 \frac{d\Omega}{dt}  &=&\mathcal W \frac{r}{h} \sin{(\omega+f)}/ \sin{i}  \,,
\\  \label{omega}
\frac{d\varpi}{dt}&=&-\mathcal{R}\frac{p}{eh}\cos{f}+\mathcal S\frac{p+r}{eh}\sin{f} \,.
\end{eqnarray}
We will work in first-order perturbation theory, whereby we express $\cal R$, $\cal S$ and $\cal W$ in terms of osculating orbit variables, set the orbit elements equal to their constant initial values in the right-hand side of Eqs.\ (\ref{a}) -- (\ref{omega}), and then integrate with respect to time.

\subsection{Time averaged variations in orbit elements}

We want to use the above equations to calculate the time averaged rates of change of the orbital elements of the target star, given by $\overline{dx/dt} \equiv T^{-1} \int_0^T (dx/dt) dt$, where $T$ is the longest relevant timescale, and $x$ is the element in question.   For an external star, $T$ would be the orbital period of the perturbing star, while for an internal star, it would be the period of the target star.  Assuming that the shorter period $P_S$ is much shorter than the longer period $P_L$ in each case, then it is straightforward to show that, modulo corrections of order of $P_{S}/P_{L}$,   
 
\begin{equation} 
\label{time-average}
\overline{\frac{dx}{dt}} \equiv \frac{1}{P} (1-e'^2)^{\frac{3}{2}}\int_0^{2\pi} \int_0^{2\pi} \frac{dx}{df}\frac{1}{(1+e'\cos{F})^2}dF df \,,
\end{equation}
where $P$ is the orbital period of the target star, $f$ and $F$  are the true anomalies of the target  and perturbing stars' orbits, respectively, $e'$ is the eccentricity of the perturbing star's orbit, and $dx/df = (r^2/h) dx/dt$, valid to first order in perturbation theory.

By way of illustration, we show here 
the time-averaged changes of orbital elements for the $\ell=1$ term induced by an external star [Eq.\ (\ref{EOM,in})], for the special case $i'=0$ and $\Omega'=0$:
\begin{eqnarray}
\label{a1}
\overline{\frac{da}{dt}}&=&0 \,,
\\ \label{e1}
\overline{\frac{de}{dt}}&=&\frac{15}{4}B_{\rm ext}\frac{e(1-e'^2)^{3/2}}{(1-e^2)^{5/2}}\sin{\omega} \cos{\omega} \sin^2{i} \,,
\\ \label{i1}
\overline{\frac{di}{dt}}&=& -\frac{15}{4}B_{\rm ext}\frac{(1-e'^2)^{3/2}}{(1-e^2)^{7/2}} e^2 \, \sin{\omega}\cos{\omega}\sin{i}\cos{i} \,,
\\ \label{Omega1}
\overline{\frac{d\Omega}{dt}}&=&- \frac{3}{4}B_{\rm ext}\frac{(1-e'^2)^{3/2}}{(1-e^2)^{7/2}}(1+4e^2-5e^2\cos^2{\omega})\cos{i} \,,
\\ \label{varpi1} 
\overline{\frac{d\varpi}{dt}}&=&\frac{3}{4} B_{\rm ext} \frac{(1-e'^2)^{3/2}}{(1-e^2)^{5/2}} \bigl ( 5\cos^2 \omega - 3 + 5 \cos^2 i \sin^2 \omega -\cos^2 i \bigr ) \,,
\end{eqnarray}
where $B_{\rm ext}=(2\pi/P)(m_3/M)(p/p')^3$. For arbitrary orientations $i'$ and $\Omega'$ the expressions are much more complicated.  We have also found the analogous expressions for the $\ell = 2$ and $\ell = 3 $ terms.   These are smaller than the $\ell = 1$ results by factors of $p/p'$ and $(p/p')^2$, respectively.

For an internal star, the $\ell=1$ term of Eq.~(\ref{EOM,in}) contributes no time-averaged variation of any of the elements.  The $\ell = 2$ contributions scale as $B_{\rm int} = (2\pi/P)(m_3/M)(p'/p)^2$, while the $\ell =3$ contributions are smaller by a factor of $p'/p$.  Again, the general expressions are long, so we will not display them here. 

Since the orbital energy of the target star is proportional to $1/a$, Eq.\ (\ref{a1})  simply reflects the absence of a secular energy exchange mechanism between the target and perturbing stars at first order in the perturbations.  As a side remark, Eqs.\ (\ref{e1}) and (\ref{i1}) together imply that $(1-e^2)^{1/2} \cos{i}$ is a constant, so that a decreasing inclination produces an increasing eccentricity; in planetary dynamics this is known as the Kozai mechanism \cite{kozai}. 

\section{Perturbations by a distribution of stars}
\label{sec:distribution}

\subsection{Average over orientations of perturbing stellar orbits}

With the time-averaged changes in the orbital elements due to one perturbing star in hand, we now turn to the changes caused by a distribution of perturbing stars.  We will assume a cluster of stars whose orbital orientations ($i'$, $\Omega'$, $\omega'$) are randomly distributed.  We will discuss the distributions in $a'$ and $e'$ later.  The ``orientation-average'' of a function will be defined by
\begin{eqnarray} \nonumber
\langle F \rangle&\equiv& \frac{1}{8\pi^2}\int_{0}^\pi  \sin{i'} \,di'\int_{0}^{2\pi} d\Omega' \int_{0}^{2\pi} d\omega' \, F(i', \Omega', \omega' ) \,.
\end{eqnarray}
We then find that $\langle \overline{dx/dt} \rangle = 0$ for all four  orbit elements $e$, $i$, $\Omega$ and $\varpi$, for both internal and external stars.  The reason is easy to understand: the averaging process is equivalent to smearing the perturbing stars' mass over a concentric set of spherically symmetric shells.   The target star will thus be moving in what amounts to a spherically symmetric, $1/r$ potential and its orbit elements will therefore be constant.  

But for a finite number of stars, the potential will not be perfectly spherically symmetric, even if the orientations are randomly distributed.  It is the effect of this discreteness that we wish to estimate.  We do this by calculating the root-mean-square (r.m.s.) angular average $[\langle (\overline{dx/dt})^2 \rangle ]^{1/2}$.  This will give an
 estimate of the ``noise'' induced in the orbital motion of the target star by the surrounding matter.  We will then compare this noise with the relativistic effects that we wish to measure.

We will focus on the quantity 
\begin{equation}
\langle (\overline {dh/dt})^2 \rangle \equiv \langle (\overline {di/dt})^2 \rangle+\sin^2{i} \langle (\overline {d\Omega/dt})^2 \rangle  \,,
\end{equation}
rather than the individual elements $i$ and $\Omega$; this quantity represents the r.m.s change in the direction of $\hat{\bm{h}}$, the normal to the orbital plane.   The leading contributions from internal and external stars are given by
\begin{eqnarray}
\langle (\overline {dh/dt})^2 \rangle_{\rm int} &=& \frac{3}{40}B_{\rm int}^2\frac{1+3e'^2+21e'^4}{(1-e'^2)^4} \,, 
\label{rmsinner}
\\
\langle (\overline {dh/dt})^2 \rangle_{\rm ext} &=& \frac{3}{40}B_{\rm ext}^2\frac{(1-e'^2)^3}{(1-e^2)^7} \left (1+3e^2+ \frac{17}{2} e^4 \right )  \,.
\label{rmsouter}
\end{eqnarray}  
Appendix A lists the separate r.m.s.\ orientation averages for $di/dt$ and $d\Omega/dt$ for internal and external stars, and for all $\ell \le 3$. 
For future use, we define the angular r.m.s. rate of change of the orbital orientation by $d\theta/dt \equiv \langle (\overline {dh/dt})^2 \rangle^{1/2}$.

\subsection{Average over size and shape of perturbing stellar orbits} 

 We now integrate over the semi-major axis $a'$ and eccentricity $e'$ of the perturbing stars.  We will use a distribution function of the form
${\cal N} g(a') h(e'^2) da' de'^2$,
where $\cal N$ is normalization factor, set by the condition
${\cal N} = N/{\cal I}$, where $N$ is the total number of stars in the distribution, and
\begin{equation}
{\cal I} =  \int h(e'^2) de'^2 \int  g(a') da' \,,
\end{equation}
where the  limits of integration will be determined by the limiting orbital elements for those stars.  Following MAMW, we will consider a range of parametrized models for the dependences $g(a')$ and $h(e'^2)$, and will consider clusters that contain both stars and stellar-mass black holes.

The variables $a'$ and $e'$ will be constrained by a number of considerations.  
The minimum pericenter distance $r_{\rm min}$ for any body will be given by the tidal-disruption radius for a star, and the capture radius for a black hole. This will therefore give the bound
\begin{equation}
a' (1-e') > r_{\rm min} \,.
\end{equation}
For $r_{\rm min}$ we will use the estimates
\begin{eqnarray}
r_{\rm min}^{star} &\approx& 4 \times 10^{-3} \, \left ( \frac{m}{m_\odot} \right )^{0.47} \left (\frac{M}{4 \times 10^6 M_\odot} \right )^{1/3} \, {\rm mpc} \,,
\nonumber \\
r_{\rm min}^{bh} &\approx& \frac{8GM}{c^2} \approx 1.5 \times 10^{-3} \left (\frac{M}{4 \times 10^6 M_\odot} \right ) \, {\rm mpc} \,.
\label{rmin}
\end{eqnarray}
These are derived in Appendix B.

However our analytic formulae for the r.m.s.\ orientation-averaged variations are valid only in the limits $p'/p \ll 1$ or $p/p' \ll 1$ for internal and external stars, respectively.  But since our target star is embedded inside the cluster of stars, there may well be perturbing stars that do not satisfy either constraint.   
On the other hand an encounter between the target star and another star that is too close could perturb the orbit so strongly that it will be unsuitable for any kind of relativity test.  Because we are looking only for an estimate of the statistical noise induced by the cloud of stars, we will try three approaches in order to capture the range of perturbations induced by the cluster.

{\em Integration I.}
Because Eqs.\ (\ref{rmsinner}) and (\ref{rmsouter}) are valid only in the extreme limits where the perturbing star is always far from the target star (so that the higher-order terms are suitably small), we cut out of the stellar distribution any stars that violate this constraint.  This yields the following conditions  on the allowed orbital elements of the perturbing stars:
(i) for an internal star, we demand that $r'_{\rm max} = a' (1+e')$ of the perturbing star be less than $r_{\rm min} = a(1-e)$ of the target star;  (ii) for an external star, we demand that $r'_{\rm min} = a' (1-e')$ of the perturbing star be greater than $r_{\rm max} = a(1+e)$ of the target star.  

For an internal star, we thus have the two conditions,
\begin{equation}
a' (1-e') > r_{\rm min} \,,
\qquad
a' (1+e') < a(1-e) \,.
\end{equation}
The maximum values of $e'$ and $a'$ allowed under these conditions are
\begin{equation}
e'_{\rm max,int} = \frac{a(1-e)-r_{\rm min}}{a(1-e)+r_{\rm min}} \,,
\qquad
a'_{\rm max,int} = a\frac{1-e}{1+e'}\,.
\end{equation}
For an external star, we have the two conditions
\begin{equation}
a' (1-e') > a(1+e) \,,
\qquad
    a' < a_{\rm max} \,,
\end{equation}
where $a_{\rm max}$ is the outer boundary of the cluster, chosen to be large enough that the effects of stars beyond this boundary are assumed to be negligible.  Following MAMW, we choose $a_{\rm max} = 4$ mpc.
The maximum $e'$ and minimum $a'$ allowed are thus 
\begin{equation}
e'_{\rm max,ext} = 1 - \frac{a(1+e)}{a_{\rm max}}  \,,
\qquad
a'_{\rm min,ext} = a \frac{1+e}{1-e'} \,.
\end{equation}
Thus the average of a function ${\cal F}(a', e')$ over this distribution will be given by
\begin{eqnarray}
\langle {\cal F} \rangle \equiv {\cal N} (J_1 + J_2) \,,
\end{eqnarray}
where
\begin{eqnarray}
J_1 ({\cal F}) &=&  \int_0^{{e'}^2_{\rm max,int}} h(e'^2) de'^2 \int_{r_{\rm min}/(1-e')}^{a'_{\rm max,int}}  g(a') {\cal F} (a',e') da' \,,
\nonumber \\
J_2 ({\cal F}) &=&  \int_0^{{e'}^2_{\rm max,ext}} h(e'^2) de'^2 \int_{a'_{\rm min,ext}}^{a_{\rm max}}  g(a') {\cal F} (a',e') da' \,.
\label{calF1}
\end{eqnarray}
However, instead of substituting ${\cal N} = N/{\cal I}$, we substitute
\begin{equation}
{\cal N} = N/ ({\cal I}_1 + {\cal I}_2) \,,
\end{equation}
where 
\begin{eqnarray}
{\cal I}_1 &=&  \int_0^{{e'}^2_{\rm max,int}} h(e'^2) de'^2 \int_{r_{\rm min}/(1-e')}^{a'_{\rm max,int}}  g(a')  da' \,,
\nonumber  \\
{\cal I}_2 &=&  \int_0^{{e'}^2_{\rm max,ext}} h(e'^2) de'^2 \int_{a'_{\rm min,ext}}^{a_{\rm max}}  g(a')  da' \,.
\end{eqnarray}
This amounts to assuming that all $N$ stars in the cluster happen to have orbit elements that satisfy our constraint.
Thus the average of the function ${\cal F}(a', e')$ will be given by
\begin{eqnarray}
\langle  {\cal F} \rangle  &=& N \frac{J_1({\cal F}) + J_2 ({\cal F})}{{\cal I}_1 + {\cal I}_2} \,.
\end{eqnarray}
Note that if ${\cal F} = 1$, we get  $\langle { \cal F} \rangle = N$.   

In our simple model, we are treating the stars and black holes as independent distributions, so the mean value of ${\cal F}$ can be written as a sum over the two normalized distributions,
\begin{eqnarray}
\langle {\cal F} \rangle = \langle {\cal F} \rangle_S + \langle {\cal F}\rangle_B \,,
\end{eqnarray}
where the only difference between the integrals for the distributions is the value of $r_{\rm min}$, which affects only the integrals ${J}_1$ and ${\cal I}_1$, and the number of particles, $N_S$ for stars, and $N_B$ for black holes, with $N = N_B + N_S$; for later use, we define $N_B/N_S \equiv R$.
Hence we obtain
\begin{equation}
\langle {\cal F} \rangle = \frac{N_S}{N} \frac{J_{1S}({\cal F}_S) + J_{2} ({\cal F}_S)}{{\cal I}_{1S} + {\cal I}_{2}} + \frac{N_B}{N} \frac{J_{1B}({\cal F}_B) + J_{2} ({\cal F}_B)}{{\cal I}_{1B} + {\cal I}_{2}}  \,.
\end{equation}
For the r.m.s.\ variations in $dh/dt$, we include all the higher-order terms shown in Appendix A.
 
{\em Integration II.}   Taking the ratio of the higher $\ell$ contributions to the orbit element variations  to the leading $\ell$ contribution (see Appendix A) reveals that the parameter controlling the relative size of the higher-order terms is the ratio $a'/a(1-e^2)$ for internal stars, and $a/a' (1-e'^2)$ for external stars.   Requiring each of these ratios in turn to be less than one, we repeat the integrals, but with new limits of integration given by
\begin{eqnarray}
e'_{\rm max,int} &=& 1-r_{\rm min}/a(1-e^2)  \,,
\qquad a'_{\rm max,int} = a(1-e^2) \,,
\nonumber \\
e'_{\rm max,ext} &=& (1 - a/a_{\rm max} )^{1/2} \,,
\qquad a'_{\rm min,ext} = a/(1-e'^2) \,.
\end{eqnarray}
This condition permits closer encounters than the condition imposed in Integration I.  Here as well, we include all higher-order contributions to the r.m.s. variations.

{\em Integration III}.   In an attempt to include even closer encounters between the target star and cluster stars, we adopt a fitting formula for the r.m.s.\ perturbations of the orbital plane that interpolates between the two limits of very distant internal and very distant external stars.  A simple formula that achieves this is given by
\begin{equation}
{\dot h}^2_{\rm fit}  = \frac{1}{\langle (\overline {dh/dt})^2 \rangle_{\rm int}^{-1}  + \langle (\overline {dh/dt})^2 \rangle_{\rm ext}^{-1} }\,,
\end{equation}
where we use only the lowest-order contributions to the r.m.s.\ variations, given by Eqs.\ (\ref{rmsinner}) and (\ref{rmsouter}).
In this case the average over the distributions becomes
\begin{equation}
\langle {\cal F} \rangle = \frac{N_S}{N} \frac{J_{S}({\cal F}_S) }{{\cal I}_{S} } + \frac{N_B}{N} \frac{J_{B}({\cal F}_B)}{{\cal I}_{B} }  \,,
\end{equation}
where the integrals now take the form
\begin{equation}
J({\cal F}) =  \int_0^{(1-r_{\rm min}/a)^2} h(e'^2) de'^2 \int_{r_{\rm min}/(1-e')}^{a_{\rm max}}  g(a') {\cal F} (a',e') da'  \,,
\end{equation}
with ${\cal I} = {J}(1)$, thereby including the full distribution of stars.

\begin{table}[t]
\begin{center}
\begin{tabular}{crrrrrrcrrrrr}
\hline
Model&$\gamma$ & $\beta$ & $M_\star$ & $\cal R$ & $N$ 
&&Model&$\gamma$ & $\beta$ & $M_\star$ & $\cal R$ & $N$ \\
\hline

1&0 & -1  & 10 & 0   & 159  &&9& 2 & 0   & 10  & 1    & 7    \\
2&0 & -1  & 10 & 1    & 29    &&10& 2 & 0   & 30  & 0    & 119   \\

3&1 & -1  & 10 & 0    & 119  &&11& 2 & 0   & 30  & 1    & 21    \\
4&1 & -1  & 10 & 1  & 21  &&12& 2 & 0   & 100 & 0   & 400   \\

5&1 & 0 & 30 & 0 & 209 &&13&2 & 0   & 100 & 1    & 72  \\
6& 1 & 0 & 30 & 1 & 43 &&14&2 & 0.5 & 100 & 0   & 400\\

7&2 & -1  & 30 & 0    & 119   &&15&2 & 0.5 & 100 & 1   & 72    \\
8&2 & -1  & 30 & 1    & 21  \\

\hline
\end{tabular}
\caption{ Parameters of the distributions}
 \label{tbl-1}
\end{center}
\end{table}

\section{Numerical results}
\label{sec:numerical}

In order to compare our analytic estimates with the results of the N-body simulations of MAMW, we will adopt as far as possible the same model assumptions.  We parametrize the distribution functions $g(a')$ and $h(e'^2)$ according to $g(a')=a'^{2-\gamma}$, and $h(e'^2) = (1-e'^2)^{-\beta}$, where $\gamma$ ranges from 0 to 2, and $\beta$ ranges from -1 to 0.5.  The values $(\gamma, \, \beta) = (2,0)$ correspond to a mass segregated distribution with isotropic velocity dispersion.  We will chose $a'_{\rm max} = 4$ mpc, arguing that the perturbing effect of the cluster outside this radius is negligible by virtue of the increasing distance from the target star and the more effective ``spherical symmetry'' of the mass distribution.  We will assume that the cluster contains stars of mass $1M_\odot$ and black holes of mass $10M_\odot$, and will consider values of the ratio of the number of black holes to the number of stars to be $R=0$ and $R=1$ (MAMW also consider the ratio $R=0.1$).  
The main difference between stars and black holes in our integrals is the factor $m_3^2$, so there will simply be a relative factor of 100 between the black hole contribution and the stellar contribution, apart from the small effect of the difference in $r_{\rm min}$ between stars and black holes.    

Of the 22 stellar distribution models listed in Table I of MAMW, we consider only the 15 models with either $R=0$ or $R=1$; these are listed in Table \ref{tbl-1}.  While $N$ denotes the total number of objects within $4$ mpc, the parameter $M_\star$ denotes the approximate total mass within $1$ mpc, and gives an idea of the perturbing environment around a close-in target star.  

\begin{figure}[t]
\begin{center}

\includegraphics[width=5in,angle=270]{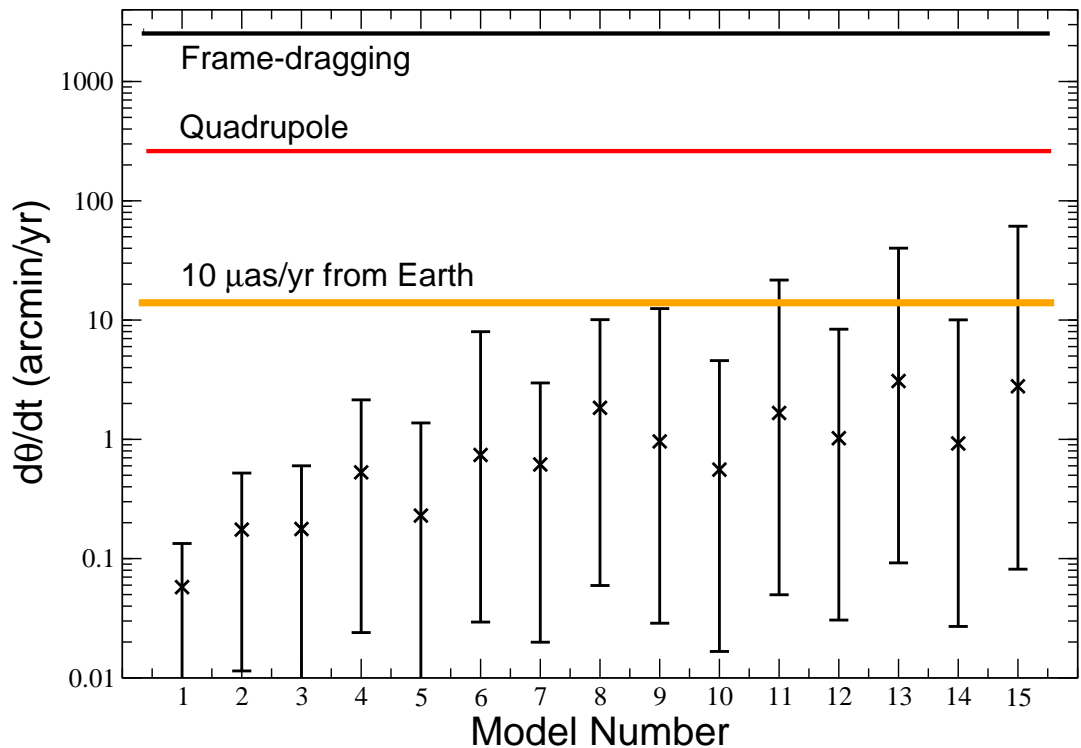}

\caption{R.m.s.\ precession $d\theta/dt =  (\langle \dot{h}^2 \rangle)^{1/2}$  for target star with $e=0.95$ and $a = 0.1$ mpc  for 15 stellar distribution models.  Symbol $\times$ denotes estimates from Integration II, and error bars indicate the range of estimates from Integrations I and III.  Amplitudes of frame-dragging, quadrupole, and astrometric displacement of 10 $\mu$as/yr are shown as in Fig.\ 1.  }
\end{center}
\label{fig2}
\end{figure}

Figure 1 shows the results for the three models 9, 11 and 12, as discussed in Sec.\ \ref{sec:intro}.
To illustrate the differences between different models of the stellar distribution, Fig.\ 2 shows the predicted precessions for a target star at $0.1$ mpc with $e = 0.95$, for all 15 model distributions.  The crosses and the error bars indicate the range of results from the three integration models.   Models with $\gamma = 0$ or  $1$ generally give smaller precessions than those with $\gamma =2$.  The latter models are more centrally condensed, and lead to larger perturbations of a close-in target star.  For the same value of $(\gamma,\, \beta, \, M_\star)$, models with equal numbers of stars and black holes $(R=1)$ lead to larger perturbations than those with pure stars ($R=0$); the former models are more ``grainy'' (smaller $N$), and so the effects are larger by roughly $N_{R=0}^{1/2}/N_{R=1}^{1/2}$.   Models 14 and 15 $(\beta = 0.5)$ have an excess of stars in highly eccentric orbits, thus leading to larger precessions.

We have tested the reliability of these estimates by carrying out a number of checks.  For Integrations I and II, we calculated the precessions first using only the lowest-order terms [Eqs.\ (\ref{h2int}), (\ref{om2int}),  (\ref{h1ext}), (\ref{om1ext})], then using those plus the first-order terms [Eqs.\ (\ref{h3int}), (\ref{om3int}),  (\ref{h2ext}), (\ref{om2ext})], and finally adding the second-order term [Eqs.\ (\ref{h3ext}), (\ref{om3ext})].  Table \ref{tbl-2} shows, for three values of $a$ and two values of $e$ the percentage change in the precession estimates for model distribution 11 caused by adding higher-order terms.  Not surprisingly Integration I suffers the smallest change, since it forces perturbing stars to be always relatively far from the target star, where the higher-order terms are relatively smaller.  Integration II suffers larger changes, as much as a factor of 2.5 for large eccentricities, but still within an order of magnitude.  

We also checked the fitting formula used in Integration III by carrying out a calculation of the r.m.s.\ precession of a target star in an eccentric orbit by a star in a circular orbit of the same semi-major axis.  The equations of motion can be formulated exactly, and the two unperturbed orbits have the same period, allowing a single timescale to be extracted.  We carried out the time averages of the perturbation equations for $di/dt$ and $d\Omega/dt$ numerically, and then carried out the averages of $(dh/dt)^2$ over $i'$ and $\Omega'$ numerically.   Finally we did a very coarse average over the phase of the circular orbit, in order to avoid the singular case where the two stars actually collide!   The result was that, for large eccentricity $e$, the fitting formula overestimates $d\theta/dt$ by about a factor of 2.

\begin{table}[t]
\begin{center}
\begin{tabular}{ccrrrr}
\hline
&&\multicolumn{2}{c}{Integration I}&\multicolumn{2}{c}{Integration II}\\
$a$ (mpc)&$e$ &1st order & 2nd order& 1st order & 2nd order\\ 
\hline
0.1&0.5& 7.5& 10.0&42.1&55.8\\
                          
&0.95&4.0&12.2&49.7&257.8\\
                        
0.5&0.5&7.2&9.4&41.9&54.3\\
                         
&0.95&3.8&12.2&50.3&255.7\\
                          
1.0&0.5&6.4&8.5&41.3&53.1\\
                         
&0.95&2.8&12.7&48.9&251.7\\

\hline
\end{tabular}
\caption{ Percentage change in $d\theta/dt$ from adding higher order terms in Integrations I and II.}
 \label{tbl-2}
\end{center}
\end{table}

\section{Conclusions}
\label{sec:conclusions}

We have used analytic orbital perturbation theory to investigate the rate of precession of the orbital plane of a target star orbiting the galactic center black hole SgrA* induced by perturbations due to other stars in the central cluster.  We found that, although the results have a wide spread, they compare well with the distribution of precessions obtained using $N$-body simulations.   One feature not included in our analysis is the fact that orbital planes in a real cluster are not randomly distributed, but become somewhat correlated over the long-term evolution of the cluster.  Whether these correlations are large enough to have a significant effect on our estimates is an open question.  Within our assumptions, however, we find a range of possible models for the cluster of objects within the central 4 mpc of the black hole in which it may still be possible to detect relativistic precessions of the orbital planes at the $10 \, \mu$as/yr level.

\ack
This work was supported in part by the National Science Foundation, Grant Nos.\ PHY 06--52448 and 09--65133, the National Aeronautics and Space Administration, Grant No.\
NNG-06GI60G, and the Centre National de la Recherche Scientifique, Programme Internationale de la Coop\'eration Scientifique (CNRS-PICS), Grant No. 4396.  LS thanks the Institut d'Astrophysique de Paris for its hospitality during the completion of this work.

\section*{Appendix A: Compendium of r.m.s. orbital perturbations}

Here we list the r.m.s.\ perturbations in $i$ and $\Omega$ for different values of $\ell$ for both internal and external stars.  It turns out that cross terms between different $\ell$ values vanish.

\noindent
Internal: Lowest order ($\ell =2$)
\begin{eqnarray}
\langle (\overline{\frac{di}{dt}})^2 \rangle_{\rm int} &=&
\frac{3}{80}B_{\rm int}^2\frac{1+3e'^2+21e'^4}{(1-e'^2)^4} \,, 
\label{h2int}\\
\langle (\overline{\frac{d\Omega}{dt}})^2 \rangle_{\rm int} &=&
\frac{3}{80}B_{\rm int}^2\frac{1+3e'^2+21e'^4}{(1-e'^2)^4} \frac{1}{\sin^{2} i} \,,
\label{om2int}
\end{eqnarray}
\noindent
Internal: First order ($\ell = 3$)
\begin{eqnarray}
\langle (\overline{\frac{di}{dt}})^2 \rangle_{\rm int} &=&
\frac{75}{7168} B_{\rm int}^2 \left ( \frac{p'}{p} \right )^2
\frac{e^2 e'^2 (6+9e'^2+34e'^4) }{(1-e'^2)^6}  (5 + 12 \cos^2 \omega )
\,, 
\label{h3int}\\
\langle (\overline{\frac{d\Omega}{dt}})^2 \rangle_{\rm int} &=&
\frac{75}{7168} B_{\rm int}^2 \left ( \frac{p'}{p} \right )^2
\frac{e^2 e'^2  (6+9e'^2+34e'^4) }{(1-e'^2)^6}   \frac{ (5 + 12 \sin^2 \omega )}{\sin^2 i} \,,
\label{om3int}
\end{eqnarray}

\noindent
External: Lowest 0rder ($\ell = 1$)
\begin{eqnarray}
\langle (\overline{\frac{di}{dt}})^2 \rangle_{\rm ext} &=&
\frac{3}{80} B_{\rm ext}^2  
\frac{(1-e'^2)^3}{(1-e^2)^7} (C_1 + D_1 \cos^2 \omega )
\,, 
\label{h1ext}\\
\langle (\overline{\frac{d\Omega}{dt}})^2 \rangle_{\rm ext} &=&
\frac{3}{80} B_{\rm ext}^2 
\frac{(1-e'^2)^3}{(1-e^2)^7} \frac{ (C_1 + D_1 \sin^2 \omega )}{\sin^2 i} \,,
\label{om1ext}
\end{eqnarray}

\noindent
External: First order ($\ell = 2$)
\begin{eqnarray}
\langle (\overline{\frac{di}{dt}})^2 \rangle_{\rm ext} &=&
\frac{225}{3584} B_{\rm ext}^2 \left ( \frac{p}{p'} \right )^2 
\frac{e^2 (1-e'^2)^3}{(1-e^2)^9} (C_2 + D_2 \cos^2 \omega )
\,, 
\label{h2ext}\\
\langle (\overline{\frac{d\Omega}{dt}})^2 \rangle_{\rm ext} &=&
\frac{225}{3584} B_{\rm ext}^2 \left ( \frac{p}{p'} \right )^2 
\frac{e^2 (1-e'^2)^3}{(1-e^2)^9} \frac{ (C_2 + D_2 \sin^2 \omega )}{\sin^2 i} \,,
\label{om2ext}
\end{eqnarray}

\noindent
External: Second order ($\ell = 3$)
\begin{eqnarray}
\langle (\overline{\frac{di}{dt}})^2 \rangle_{\rm ext} &=&
\frac{45}{4096} B_{\rm ext}^2 \left ( \frac{p}{p'} \right )^4 
\frac{(1-e'^2)^3  }{(1-e^2)^{11}}\left (1+3e'^2 +\frac{7}{2}e'^4 \right ) 
\nonumber \\
&& \quad \quad \times
 ( C_3 + D_3 \cos^2 \omega  ) \,, 
\label{h3ext} \\
\langle (\overline{\frac{d\Omega}{dt}})^2 \rangle_{\rm ext} &=&
\frac{45}{4096} B_{\rm ext}^2 \left ( \frac{p}{p'} \right )^4 
\frac{(1-e'^2)^3 }{(1-e^2)^{11}} \left (1+3e'^2 +\frac{7}{2}e'^4 \right ) 
\sin^{-2} i 
\nonumber \\
&& \quad \quad \times
 ( C_3 + D_3 \sin^2 \omega  )
 \,.
 \label{om3ext}
\end{eqnarray}
where 
\begin{eqnarray}
C_1 & = (1-e^2)^2\,,  \quad \quad \quad \quad  &D_1 = 5e^2 (2 + 3e^2) \,,
\nonumber \\
C_2 &= 5(1-e^2)^2  \,,  \quad \quad \quad \quad  &D_2  = (4+3e^2)(3+11e^2) \,,
\nonumber \\
C_3 &= (1-e^2)^2 (2+3e^2+44e^4) \,, \quad 
&D_3 = 21e^2 (2+e^2)(1+5e^2 +8e^4) \,.
\end{eqnarray}

\section*{Appendix B: Minimum distance for a stellar or black-hole orbit}

A star that approaches too close to the black hole will be tidally disrupted and be removed from the stellar distribution.  An estimate of this distance is given by the ``Roche radius'', $r_{\rm Roche} \approx R (2M/m)^{1/3}$, where $R$ is the radius of the star, and $M$ and $m$ are the black-hole and stellar masses, respectively.  For a solar-type star, the radius $R$ may be estimated using the empirical formula $R \approx R_{\odot} (m/m_{\odot} )^{0.8}$.  Thus we obtain $r_{\rm min}^{star} \approx R_{\odot} (m/m_{\odot} )^{0.47} (2M/m_\odot)^{1/3}$.  Putting in numbers gives the first of Eqs.\ (\ref{rmin}).  

A stellar-mass black hole will not be tidally disrupted, but can be captured directly if its energy and angular momentum are such that there will be no turning point in its radial motion.   For equatorial orbits in the Kerr geometry (in Boyer-Lindquist coordinates), the equation of radial motion has the form $(dr/d\tau)^2 = \tilde{E}^2 - V(r)$, where $\tau$ is $c \times$ proper time, $\tilde{E}$ is the energy per unit $mc^2$ of the body, and
\begin{equation}
V(r) = 1- \frac{2\tilde{M}}{r} + \frac{a^2}{r^2} + \frac{\beta}{r^2}
- \frac{2\tilde{M} \alpha^2}{r^3} \,,
\end{equation}
where $\tilde{M} = GM/c^2$, $a = J/Mc$, $\beta = \tilde{L}_z^2 - a^2 \tilde{E}^2$, and $\alpha = \tilde{L}_z - a\tilde{E}$, where $J$ is the angular momentum of the central black hole and $\tilde{L}_z$ is the angular momentum per unit $mc$ of the orbiting black hole.   The critical angular momentum for capture is given by that value such that the turning point occurs at the unstable peak of $V(r)$.  Since the orbiting stars and black holes are in non-relativistic orbits, we can set $\tilde{E} \approx 1$.  Under these conditions, it is straightforward to show that 
\begin{equation}
(\tilde{L}_z)_c = \pm 2\tilde{M} \left (1 + \sqrt{1 \mp a/\tilde{M}} \right ) \,,
\end{equation}
where the upper (lower) sign corresponds to prograde (retrograde) orbits.  For $a/\tilde{M} =1$, the critical angular momenta are $2\tilde{M}$ and $-2(1+\sqrt{2}) \tilde{M}$.  Converting to the language of orbital elements, where $L_z^2 =m^2 GMa(1-e^2)$, we find in the large $e$ limit, $L_z^2 \approx 2m^2 GMr_p$ where $r_p$ is the pericenter distance.  The result is that
\begin{equation}
r_{\rm min}^{bh} \approx 2\tilde{M} \left (1 + \sqrt{1 \mp a/\tilde{M}} \right )^2 \,.
\end{equation}
This ranges from $2GM/c^2$ to $11.6 GM/c^2$ for $a/\tilde M=1$
and is $8GM/c^2$ for $a=0$ (Schwarzschild).   We adopt the latter value as a suitable estimate; inserting numbers gives the second of Eqs.\ (\ref{rmin}).

\section*{Appendix C: Effects of tidal deformations}

Even if stars survive tidal disruption on passing very close to the MBH at pericenter, they will be tidally distorted, and these distortions can affect their orbits.  However, we argue that, for the stellar orbits of interest, these effects are negligible.  For example, the rate of pericenter advance due to tidal distortions is given by (Eq.\ (12.31) of \cite{tegp})
\begin{equation}
\frac{d\omega}{dt} = \frac{30 \pi}{P} k_2 \frac{M}{m} \left ( \frac{R}{a} \right )^5 \frac{1 + 3e^2/2 + e^4/8}{(1-e^2)^5} \,,
\end{equation}
where $k_2$ is the so-called ``apsidal constant'' of the star, a dimensionless measure of how centrally condensed it is.  Inserting $R = R_\odot (m/m_\odot)^{0.8}$
, we obtain
\begin{equation}
\frac{d\omega}{dt} = 0.04 \left( \frac{k_2}{10^{-2}} \right )
\left ( \frac{m}{m_\odot} \right )^{3} 
\left ( \frac{0.1 \, {\rm mpc}}{a} \right )^{13/2} 
\left ( \frac{0.05}{1-e} \right )^5  \, {\rm arcmin/yr} \,.
\end{equation}
The variations in $i$ and $\Omega$ scale in exactly the same way, but are further suppressed by the sine of the angle by which the tidal bulge points out of the orbital plane, resulting from the rotation of the star coupled with molecular viscosity, leading to a lag between the radial direction and the tidal bulge.  This angle is expected to be very small.   Thus we can conclude that, as far as perturbations of the orbital planes are concerned, tidal distortions will not be important.

\end{document}